\documentclass[a4paper]{article}
\usepackage[margin=1.2in]{geometry}
\usepackage{amssymb,amsmath, hyperref, cleveref, makecell, epsfig, amsthm, listings, tikz, subfig, subcaption, graphicx, tikz,dsfont}
\usepackage{algorithm, algpseudocode, algorithmicx, algcompatible}
\usepackage[square,numbers,sort]{natbib}
\usepackage{doi}

\usepackage{url}

\usepackage{breakurl}

\DeclareUnicodeCharacter{2212}{\textendash}

\title{A nested MLMC framework for efficient simulations on FPGAs}
\author{Irina-Beatrice Haas\thanks{irina-beatrice.haas@maths.ox.ac.uk} \and Michael B. Giles \thanks{mike.giles@maths.ox.ac.uk}}
\date{February 2025}

\hypersetup{
	pdfauthor={Irina-Beatrice Haas, Michael B. Giles},
	pdftitle={A nested MLMC framework for efficient simulations on FPGAs},
	pdfsubject={A nested MLMC framework for efficient simulations on FPGAs},
	pdfkeywords={Multilevel Monte Carlo, nested MLMC, FPGAs},
}

\begin{document}

\maketitle

\begin{abstract}
    Multilevel Monte Carlo (MLMC) reduces the total computational cost of financial option pricing by combining SDE approximations with multiple resolutions. This paper explores a further avenue for reducing cost and improving power efficiency through the use of low precision calculations on configurable hardware devices such as Field-Programmable Gate Arrays (FPGAs). We propose a new framework that exploits approximate random variables and fixed-point operations with optimised precision to generate most SDE paths with a lower cost and reduce the overall cost of the MLMC framework. We first discuss several methods for the cheap generation of approximate random Normal increments. To set the bit-width of variables in the path generation we then propose a rounding error model and optimise the precision of all variables on each MLMC level. With these key improvements, our proposed framework offers higher computational savings than the existing mixed-precision MLMC frameworks.
\end{abstract}

\section{Introduction}\label{sec:intro}

In computational finance, Monte Carlo simulations are used extensively to estimate the expected value of financial payoffs based on the solution of stochastic differential equations (SDEs) which model the evolution of stock prices, interest rates, exchange rates and other quantities \cite{glasserman04}.  Monte Carlo methods are very general and flexible, but for high accuracy it requires generating a large number of costly SDE path approximations, which has motivated research into a number of variance reduction or, equivalently, cost reduction techniques. One such method is
Multilevel Monte Carlo (MLMC), which was proposed in \cite{GILES2008} and was adapted for various applications that are summarised in \cite{Giles_overview17} and successfully combined with other methods such as quasi-Monte Carlo methods. The main idea of MLMC is to approximate the payoff using different time stepping resolutions when numerically solving the underlying SDE and to generate an optimal number of samples on each level, such that the overall computational cost is minimised subject to the desired bound on the variance. %, such that the total computational cost is minimised. 
The computational savings come from the fact that most samples are computed on the coarser levels and hence are less expensive while only a few samples from the finest levels are required \cite{GILES2008}.

Among the directions in which the computational cost 
of MLMC methods could further be reduced, an important avenue is the use of lower precision calculations, especially for the first Monte Carlo levels where the targeted accuracy is relatively low. 
 An overview of the research on mixed precision for the standard Monte Carlo (MC) framework is provided in \cite{ChowMixedPrecisionStandardMC} but only a few references study the potential of low precision computation in the MLMC framework \cite{Rounding_error_oliver}. To the best of our knowledge, the only MLMC framework with customised precision in the literature is \cite{brugger2014mixed}, but they use a uniform precision for all operations on each Monte Carlo level instead of optimising 
 the precision of each intermediary variable to reduce as much as possible the cost of path generation.
 
An important motivation for an MLMC framework with variable precision would be performing the low precision computations on reconfigurable hardware devices such as Field Programmable Gate Arrays (FPGAs). FPGAs contain customizable logic blocks and connectors that make it easy to adapt the digital circuit architecture for a specific application, leading to a highly parallel and optimised implementation. Therefore they are successfully exploited in applications that require high speed and have high computational workload, such as signal processing \cite{woods2008fpga}, and real time applications like high frequency trading \cite{HFT1,HFT2}. That is why a number of previous works in hardware architecture design implemented the MLMC algorithm to price financial options using FPGAs as accelerators, which resulted in improved speed and power efficiency compared to full CPU architectures \cite{Schryver2013AMM}. The paper \cite{lindsey2016domain} also proposed 
a Domain Specific Language to automate the configuration of FPGAs for this specific application. However, only \cite{brugger2014mixed} proposed a heuristic to reduce the precision in calculations.

In addition, all aforementioned works considered that the random number generation (RNG) is performed in single or double precision. Yet in most cases an important portion of the workload in the overall MLMC simulation comes from the RNG and in \cite{brugger2014mixed} this limited the total computational savings.
To reduce the cost of MLMC simulations in particular those based on the Geometric Brownian Motion (GBM), \cite{approximateICDF_Oliver, NestedOliver} have proposed to use approximate random numbers that are generated by applying an approximation of the inverse CDF to uniform random numbers. In \cite{NestedOliver}, the authors proposed a way to integrate these lower precision random variables into a \textit{nested} MLMC framework and completed a numerical analysis to bound the resulting error at each MC level by a product of the time step and the error in the random number approximation. The same authors show in \cite{approximateICDF_Oliver} that using approximate random variables reduces the cost of path generation by a factor 7.

In this paper we propose a nested MLMC framework that combines the use of approximate random normal variables and lower precision calculations to reduce the computational cost of MLMC even further than \cite{brugger2014mixed,NestedOliver}. We illustrate the efficiency of our framework in Matlab, after making several assumptions on the cost of operations and size of the errors that we carefully justify. We focus on the case of GBM and use the approximate RNG methods presented in \cite{approximateICDF_Oliver} as well as a new slightly modified method that combines CDF inversion and the central limit theorem. To choose the precision of the variables in the low precision path generation, we introduce a novel method to optimise the bit-widths. This optimisation is performed before the main path generation loop is executed and is based on a linear model of the payoff error  
due to rounding when computing in low precision. The error model relies on algorithmic differentiation in a similar manner to \cite{unifying-bwoptim,bitwidth-AD,ADAPT}. The bit-width optimisation procedure can be performed off-line, so this stage can be excluded from the on-line time complexity of our framework. The user specified desired accuracy is then enforced by calculating on-line the number of samples that need to be generated.

In terms of hardware design, we suggest implementing the low precision path generation on FPGAs and the full-precision ones on a CPU or GPU. 
The FPGA offers enough flexibility to define a separate bit-width for every variable in the low precision path generation, and can be reconfigured periodically to update the bit-widths when the market parameters have changed considerably.

The paper is organized as follows : \Cref{sec:MLMC} introduces MLMC and nested MLMC to make clear the estimator that is implemented in our framework. Then in \Cref{sec:RNG} we detail the methods that could be used to obtain approximate random normally distributed numbers very cheaply for the low precision path generation. In \Cref{sec:error_model} and \Cref{sec:costModel} we propose an error model and a cost model (resp.) that we then use to formulate the optimisation problem that is solved to obtain the optimal bit-widths of fixed point variables in \Cref{sec:optimisation}. Finally we summarise our results and future directions in \Cref{sec:conclusion}.

\section{Multilevel Monte Carlo and nested Multilevel Monte Carlo} \label{sec:MLMC}

In this paper we want to estimate the expectation of some payoff functional $P$ that is computed on an underlying asset price $S_t$ that follows the SDE
\begin{equation}
{\rm d}S_t = a(S_t,t)\, {\rm d}t + b(S_t,t)\, {\rm d}W_t.
\end{equation}
The solution of the SDE is approximated with the Euler-Maruyama scheme. 
Letting $S_i$
denote the approximation at time $t_i\triangleq ih$, where $h$ is the size of each time step,
the approximate solution is given by 
\begin{equation}\label{eq:GBM_sde}
    S_{i+1} = S_{i} + a(S_i,t_i)h +  b(S_i,t_i) \sqrt{h}\, Z_i
\end{equation}
where $Z_i$ is a random increment that follows the normal distribution $\mathcal{N}(0,1)$.

\subsection{Multilevel Monte Carlo}

The idea of the MLMC method is to use different levels of time discretisation in order to split the computational work such that the total cost is minimised and the desired overall accuracy is achieved. 

We use $L+1$ levels that we denote by the index $\ell\in\{0,1, \ldots, L\}$. For each level $\ell$ we define an approximate payoff $P_\ell$ computed in full-precision with a time step of size $h_\ell = 2^{-\ell} T$. For $L$ sufficiently large we have the weak convergence $\mathbb{E}[P_L] \approx \mathbb{E}[P]$. The main idea of MLMC is to decompose $\mathbb{E}[P_L]$ as follows
\begin{equation} \label{classic_MLMC_iden}
    \mathbb{E}[P_L]=\sum_{\ell=0}^L \mathbb{E}[P_\ell-P_{\ell-1}]
\end{equation}
with the convention $P_{-1}=0$. In each sample of $\Delta P_\ell \triangleq P_\ell-P_{\ell-1}$ both terms are computed using the same Brownian motion \cite{Giles_overview17,NestedOliver}, that is to say that the fine path uses the approximate solution to the SDE from the Euler-Maruyama scheme with $2^\ell$ time-steps :
\begin{equation}
    S^{f}_{i+1} = S^{f}_{i} + a(S^{f}_{i}, t_i) h + b(S^{f}_{i},t_i) \, \sqrt{h}\, Z_i
\end{equation}
and the coarser path is computed as
\begin{equation}
    S^{c}_{i+1} = S^{c}_{i} + a(S^{c}_{2\lfloor i/2\rfloor}, t_{2\lfloor i/2\rfloor})\, h + b(S^{c}_{2\lfloor i/2\rfloor},t_{2\lfloor i/2\rfloor})\, \sqrt{h} \, Z_i,
\end{equation}
so that when $i$ is even, $S^{c}_{i+1}$ and $S^{c}_{i+2}$ are both computed using the drift and volatility evaluated based on $S^{c}_{i}$, $t_i$.
For each level the expectation $\mathbb{E}[\Delta P_\ell]$ is approximated using a standard Monte Carlo estimator with $N_\ell$ samples. 
Then defining % noting
$C_\ell$ and $V_\ell$ to be the cost and variance of a sample of $\Delta P_\ell$, the overall cost of the estimation and the overall variance are 
\begin{align}
    Cost &= \sum_{\ell=0}^L N_\ell C_\ell \\
    Variance &= \sum_{\ell=0}^L N_\ell^{-1} V_\ell.
\end{align}
Using a Lagrange multiplier $\lambda\in \mathbb{R}$ and treating the number of samples $N_{\ell}$ as real variables, we then minimise the cost under the constraint $Variance = \varepsilon^2$. We obtain that the number of optimal samples on each level is $N_\ell = \lambda \sqrt{V_\ell/C_\ell}$, where $\lambda = \varepsilon^{-2} \sum_{\ell=0}^L \sqrt{V_\ell C_\ell}$ and, ignoring the small increase in the cost when $N_\ell$ are rounded up to integers, the overall cost of the MLMC estimator is
\begin{equation}\label{total_cost_MLMC}
    Cost_{MLMC}=\varepsilon^{-2} \left( \sum_{\ell=0}^L \sqrt{V_\ell C_\ell}\right)^2.
\end{equation}
For comparison, noting $V=\mathbb{V}[\hat{P}]$ and $C$ the variance and cost in the standard Monte Carlo estimator the overall cost would be $\varepsilon^{-2} VC$. As  \cite{Giles_overview17} shows, if the factor $V_\ell C_\ell$ decreases (resp. increases) with level then the total cost of MLMC is approximately $\varepsilon^{-2} V_0 C_0$ (resp. $\varepsilon^{-2} V_L C_L$) so it is smaller than the standard Monte Carlo cost by a factor $C_0/C_L$ (resp. $V_L/V_0$). Since we know that the cost of a sample increases with level and for Lipschitz payoffs for the Euler-Maruyama scheme the variance $V_\ell$ decreases exponentially with level, the former leads to the MLMC estimation being cheaper than the standard Monte Carlo estimation.

\subsection{Nested Multilevel Monte Carlo}

In the nested framework we further decompose each term from the sum \eqref{classic_MLMC_iden} such that each of the full precision expectations are decomposed into a low precision estimate and a correction term. This formulation was already used in \cite{NestedOliver} and ensures that the expectations rigorously cancel out.
Formally, we split each level expectation $\mathbb{E}[\Delta P_\ell]$ as follows :
\begin{equation} \label{chap:nested_exp}
    \mathbb{E}[P_L] = \sum_{\ell=0}^L  \mathbb{E}[\widetilde{\Delta P}_\ell] + \mathbb{E}[\Delta P_\ell - \widetilde{\Delta P}_\ell].
\end{equation}
We use tildes to denote the variables that are computed in low precision.
Again each expectation is obtained with a standard Monte Carlo estimator using $\Tilde{N}_\ell$ samples for $\mathbb{E}[\widetilde{\Delta P}_\ell]$ and $N^{\Delta}_\ell$ samples for $\mathbb{E}[\Delta P_\ell - \widetilde{\Delta P}_\ell]$. We note $\Tilde{C}_\ell$ the cost of a sample of $\widetilde{\Delta P}_\ell$ and $C^{\Delta}_\ell$ the cost of a sample of $\Delta P_\ell - \widetilde{\Delta P}_\ell$, and similarly $\Tilde{V}_\ell = \mathbb{V}[\widetilde{\Delta P}_\ell]$ and $V^{\Delta}_\ell = \mathbb{V}[\Delta P_\ell-\widetilde{\Delta P}_\ell]$.
Then the total computational cost and the total variance of the nested estimator of the output is
\begin{align}
    Cost &= \sum_{\ell=0}^L \Tilde{N}_\ell \Tilde{C}_\ell + N^{\Delta}_\ell C^{\Delta}_\ell \label{eq:cost_nested}\\
    Variance &= \sum_{\ell=0}^L \Tilde{N}_\ell^{-1} \Tilde{V}_\ell + (N^{\Delta}_\ell)^{-1} V^{\Delta}_\ell. \label{eq:var_nested}
\end{align}
Suppose we would like the overall variance to be smaller than $\varepsilon^2$, then similarly to the standard MLMC case, using a Lagrange multiplier $\lambda_M \in \mathbb{R}$ gives $N^{\Delta}_\ell = \lambda_M \sqrt{V^{\Delta}_\ell/C^{\Delta}_\ell}$ and $ \Tilde{N}_\ell = \lambda_M \sqrt{\Tilde{V}_\ell/\Tilde{C}_\ell}$ for all levels $\ell$. Plugging these expressions back in $Variance = \varepsilon^2$ gives the total cost
\begin{equation}
    Cost_{nested} = \varepsilon^{-2} \left( \sum_{\ell=0}^L \sqrt{\Tilde{V}_\ell\Tilde{C}_\ell}+\sqrt{V^{\Delta}_\ell C^{\Delta}_\ell} \right)^2.
\end{equation}
As \cite{NestedOliver} shows, roughly speaking, if $V^{\Delta}_\ell/\Tilde{V}_\ell \ll \Tilde{C}_\ell/C^{\Delta}_\ell \ll 1$ then the nested estimation leads to a computational saving of a factor approximately $\max_\ell \Tilde{C}_\ell / C^{\Delta}_\ell$ compared to the standard MLMC framework.

The cost of computing a sample of the correction term is $C^{\Delta}_\ell = C_\ell+ \Tilde{C}_\ell$ where $C_\ell$ is the cost of generating $\Delta P_\ell$ in full precision. 
The key point of our framework is that the constant $C_\ell$ is much larger than $\Tilde{C}_\ell$ at least on the first MC levels, which leads to important computational savings as most samples generated in MLMC are on these levels.
Moreover, in the low precision path generation, instead of using random increments that are in full precision we use approximate random normal increments $\Tilde{Z}_i$ as follows 
\begin{equation}\label{eq:GBM_sde_fi}
    \Tilde{S}_{i+1} = \Tilde{S}_{i} + a(\Tilde{S}_i,t_i) \, h + b(\Tilde{S}_i,t_i)\, \sqrt{h}  \, \Tilde{Z}_i.
\end{equation}
Ideally the cost of RNG on the FPGA would be almost negligible, so that the path generation on the FPGA is performed very cheaply.

\section{Approximate random normally distributed numbers} \label{sec:RNG}
In this section we summarise three methods that could be used to generate approximate random normal variables that are used in the low precision path generation. 
All methods below are classified as \textit{inversion methods}, because they are based on applying an approximation of the inverse normal CDF $\Phi$ to a random uniform variable $U$. The first and the third methods approximate $\Phi^{-1}$ by a Piecewise Constant (PWC) function on uniform intervals included in $[0,1]$. The second method is based on previous work by Lee et al. \cite{Cheung2007HardwareGO, Lee_segmentation} and approximates $\Phi^{-1}$ by a Piecewise Linear (PWL) function on dyadic intervals, ie. intervals that are progressively divided by 2 as we get closer to the singularities of $\Phi^{-1}$. The approximation accuracy of the first and second methods were analysed asymptotically in \cite{approximateICDF_Oliver}. Note that, in practice, we exploit the symmetry of $\Phi^{-1}$ so we only need to approximate it on $[0,1/2]$.

In order to compute the correction terms we need to be able to generate a couple of random normal increments $(Z,\Tilde{Z})$ in double and reduced precision respectively.
In all three approaches below, $Z$ is computed on the CPU either with a piecewise polynomial approximation of $\Phi^{-1}$ of degree 5 on the hierarchical interval segmentation described in \cite{Cheung2007HardwareGO}, or more directly using a math routine that applies $\Phi^{-1}$ to a vector such as in \cite{norminv_routine_intel}. Therefore for each of the following methods we only detail how to obtain $\Tilde{Z}$ and the corresponding full precision uniform variable $U$ (from which computing $Z$ is trivial).

To fix the notation, we consider that the double precision uniform variable $U$ corresponds to a $D$-bit integer that we denote by $J$ and the associated low precision uniform variable corresponds to a $d$-bit integer denoted by $j$.

\subsection{Piecewise constant approximation on uniform intervals (Method 1)} %\label{method1}
In the first approach, the inverse CDF $\Phi^{-1}$ is approximated by a constant value on uniform intervals $\mathcal{I}_j=[u_j , u_{j+1}]\subset [0,1/2]$  with $u_j = 2^{-d} j, j<2^{d-1}$. We simply construct a Look-Up-Table (LUT) of size $2^{d-1}$ containing the constant values $Z_j$ that the approximate random number takes when the input uniform variable is inside the interval $\mathcal{I}$. The leading bit of $j$ gives the sign of the normal increment, which allows us to extend the approximation of $\Phi^{-1}$ on the interval $[1/2,1]$. The next $d-1$ bits are used to pick the right value in a LUT.
Locating the interval corresponding to the input uniform variable $U$ is trivial since the integer $j$ maps to the index of the interval. Therefore the evaluation of the approximate random number is cheap once the random integer $J$ has been generated.

To determine the optimal constants $Z_j$ the mean squared error (MSE)
\begin{equation}
    \int_{u_j}^{u_{j+1}} \left(Z_j-\Phi^{-1}(u) \right)^2 du
\end{equation}
is minimised with respect to the LUT values $Z_j$. This gives that $Z_j$ is the mean of $\Phi^{-1}$ over the interval $\mathcal{I}_j$ :
\begin{equation}
    Z_j = 2^d \int_{u_j}^{u_{j+1}} \Phi^{-1}(u)du.
\end{equation}
Then the corresponding full precision uniform variable is defined as $U = 2^{- D}\left(J + \frac{1}{2}\right)$. 

The issue with this approach is that the LUT is of size $2^{d-1}$, which may not fit on the FPGA or may take up too much hardware resources, for example for $d=10$ the LUT stores $2^9=512$ values.

\subsection{Sum of several variables (Method 2)} %\label{section_meth3}
Another approach is to reduce further the precision of the approximate random variables generated with the LUT from method 1 and sum several of them to improve the statistical quality of the final approximate normal variables. For example take an integer $n$ that divides $d$ and produce an approximate random number $X^{(1)}$ from the first $d/n$ bits of $j$, then $X^{(2)}$ from the next $d/n$ bits and so on, where each $X^{(i)}$ follows approximately the distribution $\mathcal{N}(0,1/n)$.
Then $\sum_{i=1}^n X^{(i)}$ has approximately the distribution $\mathcal{N}(0,1)$.

The idea of summing several approximate random numbers was already used in \cite{Thomas2009ACO,Malik2016GaussianRN,thom14} with $n=8$, $n=4$ and $n=2$ respectively. In particular, \cite{Thomas2009ACO} used the piecewise linear approximation on dyadic intervals (see next subsection) to generate the lower precision random numbers $X^{(i)}$.

The difference between the method introduced below and the previous literature is an optimisation stage where the coefficients stored in the LUT are adapted iteratively, which improves the quality of the estimation compared to simply taking the LUT values $Z_j$ described previously. 
The second difference compared to previous work is the coupling between the low precision and full precision uniform variables that is required in our application.
We explain the method for two variables ($n=2$), as it is generalised in a trivial way. 

Take the integer $j$ and split it into two integers with bit-widths $d/2$. Both streams of lower precision variables $X^{(1)},X^{(2)}$ are computed with the same LUT of size $2^{d/2-1}$. This LUT is initialised using the method 1 and dividing the LUT values by $\sqrt{2}$. Therefore let's note $X_j$ the values in the small LUT. Using this initial LUT 
we form a larger LUT of size $2^{d}$ by computing all possible sums $\pm X_k\pm X_l$ and ordering the outputs $\Tilde{Z}_j$ in ascending order. This step defines a permutation $\pi$ such that $\pi(j)$ gives the position of the random number $\Tilde{Z}_j$ in the ordered list. Then we perform a least-squared minimisation of 
\begin{equation}
    \sum_{j=1}^{2^{d}} (Z_{\pi(j)}-\Tilde{Z}_j)^2
\end{equation}
where the $Z_{\pi(j)}$ are obtained with a LUT of size $2^{d-1}$ as in the previous subsection.
In this minimisation problem, the variables $\Tilde{Z}_j$ are considered as linear variables of the decision variables, which are the values $X_j$ from the small LUT that we want to optimise.

After this step update the permutation $\pi$ by ordering the resulting $\Tilde{Z}_j$ and repeat the least-squares optimisation. The algorithm stops when the permutation has converged.
In practice we observed that for $d=10$ and $d=12$ this process takes about 20 and 100 iterations respectively, and that the optimisation stage considerably reduces the MSE (see \Cref{fig:MSE_plot}).

Finally, generating the coupled full precision variable requires taking into account the permutation $\pi$. When the optimised LUT is used and the input integer is $j$, the output is the value $\Tilde{Z}_j$, which is an approximate of $\Phi^{-1}(2^{-d}\pi (j))$. Therefore the corresponding uniform variable used on the CPU is
\begin{equation}
    U = 2^{-d}\pi (j) + 2^{-D}\left((J-2^{D-d}j)+ \frac{1}{2}\right) = 2^{-d}(\pi (j) -j ) + 2^{-D}\left(J+\frac{1}{2}\right).
\end{equation}
The permutation $\pi$ only needs to be stored (in a permutation table of size $2^{d}$) and used on the CPU while the RNG on the FPGA is done using only the optimised LUT of size $2^{d/2-1}$.

Generalising the method to $n>2$ variables is straightforward. The only change is that the values of the small LUT are divided by $\sqrt{n}$ and $n$ low precision variables are summed to obtain the $\Tilde{Z}_j$. The coupled uniform variable used on the CPU is defined with exactly the same formula as in the two variables case.

\subsection{Piecewise linear approximation on dyadic intervals \linebreak (Method 3)}
Another way to reduce the size of the LUT is to use dyadic intervals as suggested in \cite{Cheung2007HardwareGO,approximateICDF_Oliver}. Again we consider that the leading bit of $j$ is a sign bit used to flip the sign of an approximate normal variable when $U>\frac{1}{2}$ and the next bits are used for the approximation of $\Phi^{-1}(U)$.
The approximate random number corresponding to integer $j$ is defined as $\Bar{Z}_j = a + bj$, with a separate pair $(a, b)$ for each interval $\mathcal{I}'_i=[\![ 2^{i-1}, 2^{i}-1]\!]$, where $i$ is the leading non-zero bit of the integer $j$. The coefficients are stored in a LUT of size $d - 1$ and the values $(a, b)$ are again obtained by minimising the MSE. A simple calculation shows that
\begin{equation} \label{mse_dyadic}
    \int_{u_j}^{u_{j+1}} (\bar{Z}_j - \Phi^{-1}(u))^2 du = 2^{−d} (\bar{Z}_j - Z_j )^2 + \int_{u_j}^{u_{j+1}} (Z_j - \Phi^{-1}(u))^2 du
\end{equation}
where $Z_j$ is defined as in method 1. Therefore to calculate the pairs $(a, b)$ we only need to minimise
\begin{equation}
    \sum_{j=1}^{2^{d-1}} (\Bar{Z}_j - Z_j )^2.
\end{equation}

\Cref{mse_dyadic} also shows that for the same value of $d$ this approximation cannot be as good as the method 1, but the motivation for method 3 is that the size of the LUT is considerably reduced.
The uniform CPU variable $U$ is defined in the same way as in method 1 and the coupling $(\Tilde{Z} , Z)$ follows naturally.

\subsection{Comparison of the three inversion methods}
\begin{figure}[!h]
    \centering
    \includegraphics[width=0.7\linewidth]{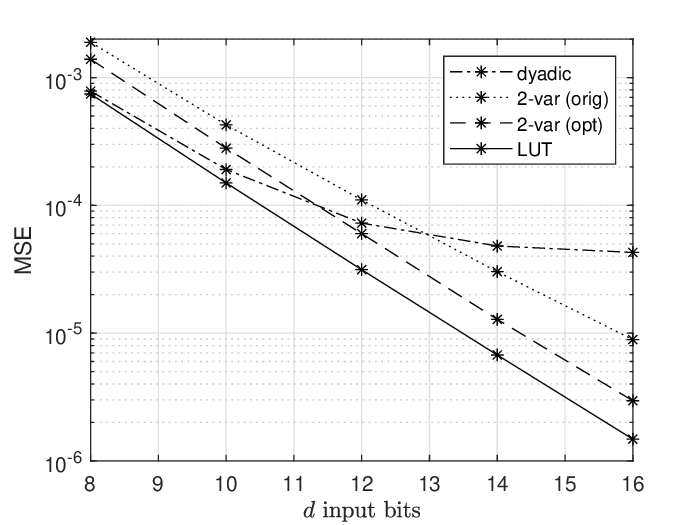}
    \caption{Mean-squared error for RNG methods 1, 2 (before and after LUT optimisation) and 3 over the bit-width $d$ of the random integer $j$.} 
    \label{fig:MSE_plot}
\end{figure}

In this section we discuss the advantages and drawbacks of the approximate RNG methods above.
We compare roughly the three approximation methods by looking at the resulting MSE (see \Cref{fig:MSE_plot}).

First note that, for the same $d$, the MSE obtained in methods 2 and 3 are necessarily larger than in method 1 (see \eqref{mse_dyadic}). Despite this, for $d=10$ the dyadic approximation is only slightly worse, which is probably because for the first two dyadic intervals there are only two points in each so $\Bar{Z}_j$ will exactly match $Z_j$. However for $d=12$ there is nearly a factor 2 difference, which illustrates that the MSE does not evolve in the same way depending on the type of intervals that is used. As illustrated by \Cref{fig:MSE_plot} and the convergence analysis from \cite{approximateICDF_Oliver},
as $d$ tends to infinity, the uniform intervals give $MSE \longrightarrow 0$ and the dyadic intervals give $MSE \longrightarrow C$, for some positive constant $C$. The latter is because with our simple dyadic segmentation, when $d$ increases by 1 the MSE is reduced only in the interval closest to 0, so the error due to the other intervals remains the same.
This is the reason why the dyadic intervals were split further into smaller uniform intervals in \cite{Lee_segmentation} as it improves the approximation of the inverse CDF. 
However their address location process is too complex for our needs, so we do not consider improving the segmentation of the dyadic method further.

Next, it is also interesting to compare methods 2 with $d$ bits against the method 1 with $d-1$ bits, since most of the values in method 2 occur in pairs ($X_i+X_j$ and $X_j+X_i$). The corresponding MSE values are very close but looking at the actual values, we saw that method 2 is slightly more accurate, probably due to the extra values of the form $X_i+X_i$.
The \Cref{fig:MSE_plot} also illustrates that both method 1 and 2 have their MSE divided by 2 each time $d$ increases by 1. This was theoretically expected for method 1 (see \cite{approximateICDF_Oliver}) and is intuitively true for method 2 after the optimisation stage as the MSE is minimised such that the $\Tilde{Z}_j$ approximate the $Z_j$ corresponding to a larger LUT, so the slope of the MSE in method 2 should be similar to that of method 1. 

Hence the double variable approach shows good accuracy for significantly reduced LUT size, with only a factor 2 increase in the MSE for the same value of $d$. The optimisation stage contributes significantly to the accuracy of method 2. This might be because the optimisation allows some values of the table to take more "extreme" values to better approximate the tail of the distribution. This is consistent with an increase in the maximal value after optimisation that we have observed. %(see \Cref{tab:error_comp}).

In conclusion, methods 2 and 3 are suitable for computing approximate random normal variables that could be used in the nested framework very cheaply and with relatively small hardware requirements. However tests on real hardware are needed to
verify the efficiency of their implementation on current FPGAs. 
Also with dyadic intervals, the MSE value cannot be arbitrarily small, which might limit the number of levels where this method is relevant.

\section{Error model based on a linear approximation} \label{sec:error_model}
In order to choose the right precision in the FPGA calculations we need to approximate the overall error caused by the accumulation of rounding errors in the low precision path simulation. In this section, focusing on the single level Monte Carlo estimator to simplify the exposition, we model the overall rounding error $P-\Tilde{P}$ and find an estimate of its variance.

\subsection{Fixed-point arithmetic and notation}
The low precision calculations would be performed on the FPGA in fixed-point arithmetic, therefore contrarily to calculating in floating-point arithmetic here the size and precision of each variable are fixed. We denote $x_i, \: i=1, \ldots, m$ the intermediary variables computed to obtain the final output $P$. As a testcase, we consider Geometric Brownian Motion (GBM) and distinguished the intermediary variables $con1, con2$ etc. whose definitions are given in \Cref{alg:GMB_code_forward}. This algorithm details how we decomposed the operations to compute each time step from \eqref{eq:GBM_sde_fi}.

\begin{algorithm}[h]
\caption{Geometric Brownian Motion path calculation (decomposed in elementary operations to show the intermediary fixed-point variables)}\label{alg:GMB_code_forward}
\begin{algorithmic}
%\State \textbf{Inputs:} interest rate $r$, volatility $\sigma$, maturity $T$, initial asset price $S_0$, number of time steps $N$
\STATE $h \gets T/N$
\STATE $con1 \gets r \times h$
\STATE $con2 \gets \sqrt{h} \times \sigma$
\STATE generate (approximate) random normal increments $\Tilde{Z}_i$ for $i=1, \ldots, N$ % this version of the algorithm generates all normals before the for loop because the same normals are used in the backward loop -- otherwise the RNG would be inside the for loop below
\FOR{$i=1, \ldots, N$}
\STATE $mul1_i \gets con2 \times \Tilde{Z}_i$
\STATE $sum1_i \gets con1 + mul1_i$
\STATE $mul2_i \gets S_i \times sum1_i$
\STATE $S_{i+1} \gets S_i + mul2_i$
\ENDFOR
\end{algorithmic}
\end{algorithm}

The fixed-point equivalent of variable $x_i$ is defined as $\Tilde{x}_i = (-1)^s 2^{e_i-d_i}n$, where $n$ is a non negative integer smaller than $2^{d_i}$. The parameters $e_i$ are called \textit{exponents} and represent the size of the variables. We determined them from running $10^6$ paths and considering the maximum absolute value that occurs for each variable. The variables $d_i$ are the \textit{bit-widths} and represent the \textit{precision} of the variable $\Tilde{x}_i$. They are the decision variables in our optimisation problem. Finally here $s$ is the value of the sign bit. For simplicity we consider that all variables are signed, therefore the total word-length of variable $\Tilde{x}_i$ is $d_i+1$. We note the rounding error on each operation as $\delta x_i = x_i - \Tilde{x}_i$ and used round-to-nearest when truncating the result of a bit-wise operation, which leads to the following bound :
\begin{equation}\label{bound_rounding_error}
    |\delta x_i | \leq 2^{e_i-d_i-1}.
\end{equation}
Without making any assumption on the distribution of the errors we then have that
\begin{equation} \label{bound_err2}
    \mathbb{E}[\delta x_i^2] = 4^{e_i-d_i-1}.
\end{equation}
If additionally we assume that the rounding errors $\delta x_i$ are uniform variables over $[-2^{e_i-d_i-1},\\ 2^{e_i-d_i-1}]$ we get the expected squared rounding error
\begin{equation} \label{bound_err}
    \mathbb{E}[\delta x_i^2] = 4^{e_i-d_i}/12.
\end{equation}
These bounds will be used below to derive a bound on the error variance $\mathbb{V}[P-\Tilde{P}]$ that depends explicitly on the bit-widths.

\subsection{Error model when all variables are truncated to fixed-point arithmetic}
Now we can approximate the error in 
the output $P$ when we approximate it by its lower precision analogue $\Tilde{P}$, which is computed in fixed-point arithmetic based on the fixed-precision intermediary variables $\Tilde{x}_i$. 
We consider that the rounding errors $\delta x_i$ are small and use a first order Taylor expansion to approximate the overall error as
\begin{equation}
    P- \Tilde{P}  \approx \sum_{i=1}^m \frac{\partial P}{\partial x_i} \delta x_i.
\end{equation} 
The partial derivatives of the payoff represent how sensitive the output is to an error in the variable $x_i$. We define the \textit{sensitivity} of the payoff $P$ to variable $x_i$ as
\begin{equation}
    \Bar{x}_i = \frac{\partial P}{\partial x_i}. % or \triangleq
\end{equation}
We use algorithmic differentiation \cite{unifying-bwoptim,bitwidth-AD,ADAPT} and $10^6$ paths to compute the sensitivities $\Bar{x}_i$ at the same time as the exponents $e_i$.
Then the variance of the overall error is
\begin{equation} \label{overall_var}
    \mathbb{V}[P- \Tilde{P}] = \sum_{i=1}^m \mathbb{V}[\Bar{x}_i\delta x_i] + 2 \sum_{i\ne j\in[1,m]} Cov\left(\Bar{x}_i\delta x_i,\Bar{x}_j\delta x_j\right).
\end{equation}

Note that in our application some inputs are random normally distributed variables, therefore the sensitivities and the rounding errors are also random. We assume that the sensitivities $\bar{x}_i$ and the rounding errors $\delta x_i$ are independent for all $i=1,\ldots, m$. It always holds that $\mathbb{V}[\Bar{x}_i \delta x_i] \leq \mathbb{E}[\Bar{x}_i^2 \delta x_i^2]$ so we bound the overall error \Cref{overall_var} in two ways.

If we assume that the individual errors $\Bar{x}_i\delta x_i$ are independent and use \eqref{bound_err} we get the \textit{optimistic} upper bound on the path error variance \eqref{overall_var} :
\begin{equation} \label{optimistic}
    \mathbb{V}[P- \Tilde{P}] \leq \frac{1}{12}\sum_{i=1}^m \mathbb{E}[\Bar{x}_i^2] \: 4^{e_i-d_i} \triangleq V_{indep}(d_1,\ldots,d_m)
\end{equation}
On the other hand, if we assume perfect correlation between errors, \\ ie.  $Cov\left(\Bar{x}_i\delta x_i,\Bar{x}_j\delta x_j\right) = \sqrt{\mathbb{V}[\Bar{x}_i\delta x_i]\mathbb{V}[\Bar{x}_j\delta x_j]}$ and use the more general bound \eqref{bound_err2} then we obtain a \textit{pessimistic} upper bound :
\begin{equation} \label{pessimistic}
    \mathbb{V}[P- \Tilde{P}] \leq \left( \sum_{i=1}^m \sqrt{\mathbb{E}[\Bar{x}_i^2]}\: 2^{e_i-d_i-1} \right)^2 \triangleq V_{corr}(d_1,\ldots,d_m).
\end{equation}
These expressions do not contain the rounding errors and depend explicitly on the bit-widths $d_i$.

\subsection{Extended error model that incorporates the approximate random numbers}
In addition if we consider that at each time step we use approximate random normal increments, we simply modify the previous bounds :
\begin{align} \label{eq:extended}
    V'_{indep}(d_1,\ldots,d_m) &= \frac{1}{12}\sum_{i=1}^{m'} \mathbb{E}[\Bar{x}_i^2]\: 4^{e_i-d_i} + \frac{1}{12}\sum_{j=1}^{N} \mathbb{E}[\Bar{Z}_j^2]\times MSE\\
    V'_{corr}(d_1,\ldots,d_m) &= \left( \sum_{i=1}^{m'} \sqrt{\mathbb{E}[\Bar{x}_i^2]}\: 2^{e_i-d_i-1} + \sum_{j=1}^{N}\sqrt{\mathbb{E}[\Bar{Z}_j^2]\times MSE} \right)^2.
\end{align}
Here $N$ is the number of time steps, $Z_j$ are the random normal increments used in the full precision path generation, $\Bar{Z}_j$ is the sensitivity $\partial P/\partial Z_j$, and $MSE =\mathbb{E}[|Z_j-\Tilde{Z}_j|^2]$ where the lower precision increment $\Tilde{Z}_j$ is obtained with one of the methods from \Cref{sec:RNG}.

Therefore to integrate both approximate random variables and fixed-point arithmetic, in practice at each level after the bit-widths for all variables (including $Z$) are optimised we can for instance choose the size of the LUT such that the term containing the MSE in \eqref{eq:extended} is smaller or equal to $V_{indep}$.

\subsection{Numerical experiments} To check the validity of our model, we performed several numerical experiments in Matlab using the Fixed-Point Designer toolbox \cite{fi_toolbox} which allows the user to specify the number of bits and the exponent of each variable.
In our tests we focused on a European vanilla call option based on GBM with interest rate $r=0.05$, volatility $\sigma=0.2$, and maturity $T=1$ for $N=1$ and $N=16$ time steps. For $N=1$ there is a single Monte Carlo level, while for $N=16$ we considered the fourth MLMC correction level. Note that in this section the low precision random increments $\Tilde{Z}_i$ were obtained only by truncating the full precision ones to $d$ bits.

\begin{figure}[h]
\centering
\begin{minipage}{.5\textwidth}
  \centering
  \includegraphics[width=1\linewidth]{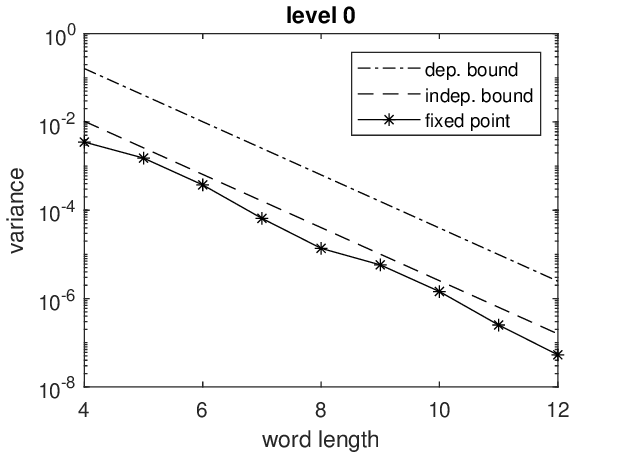}
\end{minipage}%
\begin{minipage}{.5\textwidth}
  \centering
  \includegraphics[width=1\linewidth]{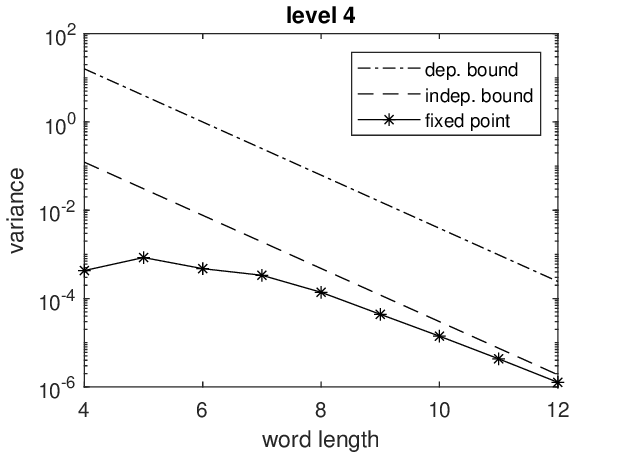}
\end{minipage}
\caption{Simulated variance of the error and variance estimates that assume independence or perfect correlation of errors, for $N = 1$ (left) and $N=16$ (right) time steps.}
\label{fig:error_model1}
\end{figure}

First of all we implemented the backward sensitivity analysis.
Then we compared the bounds \eqref{optimistic} and \eqref{pessimistic} and the variance $\mathbb{V}[P-\Tilde{P}]$ obtained from running  paths. As shown in \Cref{fig:error_model1}, where we used the same bit-width for all variables, the bounds are respected, therefore we take the expression \eqref{optimistic} to approximate the variance of the correction terms in the next sections, because this bound is tighter than \eqref{pessimistic}.

We also made several tests to check that the bound is valid including when instead of rounding the random normal variables to get the low precision normal increments we use approximate random normal variables as described in \Cref{sec:RNG}. The tests confirmed that the variance obtained numerically with approximate RNG respects the bounds $V'_{indep},V'_{corr}$. Therefore we use the bound that assumes independence of errors to approximate $V^{\Delta}_\ell$ in the overall bit-width optimisation.

\section{Cost model for the path generation}\label{sec:costModel}
In this section we define an expression for the sample costs $C^{\Delta}_\ell, \Tilde{C}_\ell$ as functions of the bit-widths $d_{i,\ell}$. This will allow us to obtain the full expression of the optimised objective as a function of known quantities and the bit-widths.

On the FPGA, if we neglect the cost of RNG, the cost $\Tilde{C}_\ell$ of computing $\widetilde{\Delta P}_\ell$ equals the sum of the cost of the elementary operations. In particular for a classical European option based on a GBM there are only additions and multiplications between two variables. 
According to the cost model from \cite{Lee2006}, letting $\mathcal{M}$ be the index set of all pairs of variables that are involved in a multiplication and $\mathcal{S}$ the analogue for additions, the cost of computing a path of $\widetilde{\Delta P}_\ell$ is 
\begin{equation} \label{cost_real}
    \Tilde{C}_\ell(d_{1,\ell},\ldots,d_{m_\ell,\ell}) = \sum_{(i,j) \in \mathcal{M}}  d_{i,\ell} d_{j,\ell} + \sum_{(i,j) \in \mathcal{S}}  \max(d_{i,\ell},d_{j,\ell}). % + \Tilde{c}
\end{equation}
However, replacing $d_{i,\ell} d_{j,\ell}$ by its upper bound $\frac{1}{2}(d^2_{i,\ell} + d^2_{j,\ell})$, and $\max(d_{i,\ell},d_{j,\ell})$ by its upper bound $d_{i,\ell}+d_{j,\ell}$ we instead use the following form for the cost :
\begin{equation} \label{cost_sq} 
    \Tilde{C}_\ell(d_{1,\ell},\ldots,d_{m_\ell,\ell}) = \frac{1}{2}\sum_{i=1}^{m_\ell} M_{i,\ell} d_{i,\ell}^2 + \sum_{i=1}^{m_\ell} M'_{i,\ell} d_{i,\ell}
\end{equation}
where $M_{i,\ell}$ (resp. $ M'_{i,\ell}$), denote the number of multiplications (resp. additions) in which the variable $x_i$ % $i$ 
is involved. %This is an over-estimation since, assuming all bit-widths are larger than 2, $d_a + d_b \leq \frac{1}{2}(d_a^2 + d_b^2)$ and $d_a \times d_b \leq \frac{1}{2}(d_a^2 + d_b^2)$.
This form of the cost model is a sum of terms which depend on only one $d_{i,\ell}$; this will be an advantage in the optimisation which is described next.

On the other hand, on the CPU most of the computational cost comes from generating a full precision random normal increment $Z_i$ at each time step ; the other operations performed to compute $\Delta P_\ell$ have a negligible cost. Hence noting $C_{RNG}$ the computational cost of one full precision normal number we get $C_\ell \approx 2^\ell C_{RNG}$.
The improvement in our framework comes from the fact that, at least on the first levels, $C_{RNG}$ is very large compared to $\Tilde{C}_\ell/N$ so using approximate random variables on the FPGA and optimal bit-widths allows one to compute only a few expensive CPU paths.

\section{Global bit-width optimisation} \label{sec:optimisation}
In the previous sections we have determined a model of the error made at each level when computing in low precision in fixed-point arithmetic 
and a model of the cost of each sample.
In this section we detail how to fix the optimal bit-widths $d_{i,\ell}$ of each variable $i$ on each level $\ell$ such that the overall cost \eqref{eq:cost_nested} is minimised and the overall variance \eqref{eq:var_nested} is under a user-specified threshold.

As mentioned in \Cref{sec:MLMC} the total cost after the number of samples in the nested framework were optimised is 
\begin{equation}
\varepsilon^{-2}
    \left(\sum_{\ell=0}^L \sqrt{\Tilde{V}_\ell \Tilde{C}_\ell}+ \sqrt{V^{\Delta}_\ell C^{\Delta}_\ell} \right)^2.
\end{equation}
We make the assumption that the variance $\Tilde{V}_\ell$ of the FPGA sample 
is approximately equal to $V_\ell \triangleq \mathbb{V}[\Delta P_\ell]$. Therefore it will be possible to compute it on the CPU before optimising the bit-widths. The variance $V^{\Delta}_\ell$ of the correction term depends on the precision used in the FPGA calculation and is approximated by $V_{indep}(d_{1,\ell}, \ldots, d_{m_\ell,\ell})$ as justified in the previous section.
Therefore the bit-widths of all variables of level $\ell$ can be optimised independently of the other levels by minimising %the level cost 
\begin{equation} \label{eq:level_cost}
    \sqrt{\Tilde{V}_\ell \Tilde{C}_\ell}+ \sqrt{V^{\Delta}_\ell C^{\Delta}_\ell}.
\end{equation}
Then the number of samples that need to be generated to determine the variance and expectation estimates can be derived analytically as shown in \Cref{sec:MLMC}.

Note that this optimisation is independent of the overall desired accuracy $\varepsilon$, therefore the bit-widths can be optimised off-line from time to time (for example once a month as the market parameters evolve slowly), which saves the time of solving the optimisation problem on-line. The on-line MLMC simulation is then similar to the classical MLMC algorithm described in \cite{GILES2008}, the only difference being that we need to approximate two expectations at each level, $\mathbb{E}[\widetilde{\Delta P}_\ell]$ and $\mathbb{E}[\Delta P_\ell - \widetilde{\Delta P}_\ell]$.

\subsection{Bit-width optimisation using a Lagrange multiplier} 
To optimise the bit-widths of all variables in the nested MLMC framework, at each level the aim is to minimise the level cost \eqref{eq:level_cost} which, using the fact that $C_\ell \gg \Tilde{C}_\ell$, is approximated as
\begin{equation}\label{eq:level_cost2}
    \sqrt{V_\ell \Tilde{C}_\ell} + \sqrt{V^{\Delta}_\ell C_\ell}.
\end{equation}
To do this, we use a Lagrange multiplier approach. To give the intuition, minimising $V^{\Delta}_\ell(d)$ subject to a fixed $\Tilde{C}_\ell(d)$ leads to the equation
\begin{equation} \label{Lagr}
    \dfrac{\partial V^{\Delta}_\ell}{\partial d_{i,\ell}} +\lambda \, \dfrac{\partial \Tilde{C}_\ell}{\partial d_{i,\ell}} = 0
\end{equation}
for a value of the Lagrange multiplier $\lambda$ which gives the desired $\Tilde{C}_\ell(d)$. Hence, the Lagrange multiplier $\lambda$ controls the trade-off between cost and variance. 
Note also that because of the form of the variance bound (\ref{optimistic}) and cost (\ref{cost_sq}), equation (\ref{Lagr}) gives a set of uncoupled nonlinear scalar equations for each pair $i, \ell$, which are easily solved to obtain $d_{i,\ell}$.

Similarly, minimising \eqref{eq:level_cost2} by equating its derivative to zero gives 
\begin{equation}
    \sqrt{C_\ell/V^{\Delta}_\ell(d)}\ \dfrac{\partial V^{\Delta}_\ell}{\partial d_{i,\ell}} +\sqrt{V_\ell/\Tilde{C}_\ell(d)}\ \dfrac{\partial \Tilde{C}_\ell}{\partial d_{i,\ell}} = 0
\end{equation}
so it is again of the form \eqref{Lagr}, where $\lambda = \sqrt{V_\ell V_\ell^{\Delta}(d)/C_\ell \Tilde{C}_\ell(d)}$ gives the optimal trade-off between cost and variance. The idea is therefore that given a guess of $\lambda$ we solve iteratively a system for the bit-widths $d_{i,\ell}$, then update the value of $\lambda$ until we reach the optimal solution of the overall problem.

\begin{figure}
    \centering
    \begin{minipage}{.48\textwidth}
        \centering
        \includegraphics[width=1\linewidth]{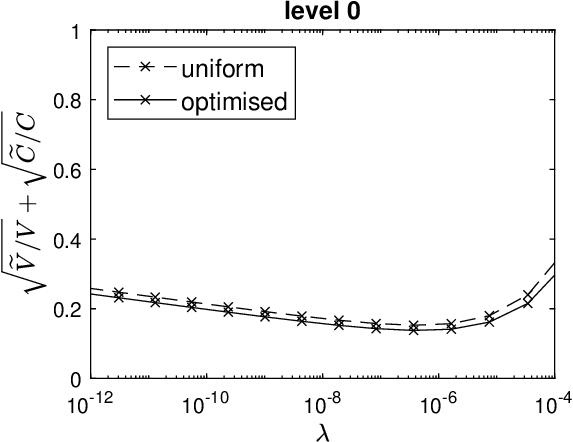}
    \end{minipage}
    \begin{minipage}{.48\textwidth}
        \centering
        \includegraphics[width=1\linewidth]{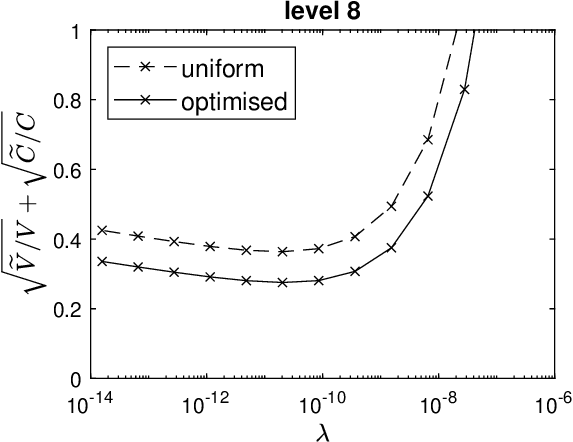}
    \end{minipage}
    \caption{$\sqrt{\Tilde{C}/C}+ \sqrt{\Tilde{V}/V}$ versus $\lambda$ for levels 0 and 8.}   \label{fig:cost_lambda}
\end{figure}

\begin{figure}[h]
\centering
\begin{minipage}{.48\textwidth}
%  \centering
  \includegraphics[width=0.9\linewidth]{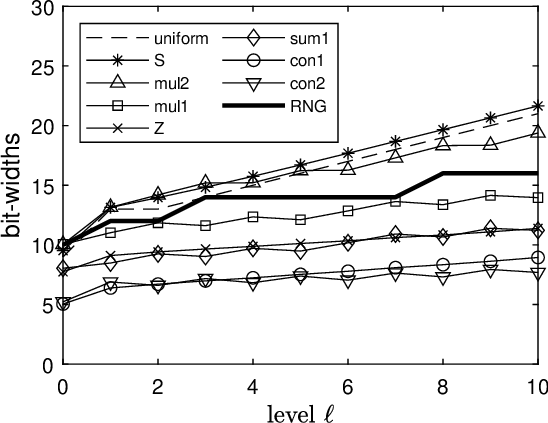}
  \captionof{figure}{Optimal bit-widths for each variable, best uniform bit-width, and required RNG accuracy, all versus level.}
  \label{fig:optimal_bw}
\end{minipage}%
\hspace{.02\textwidth}
\begin{minipage}{.48\textwidth}
%  \centering
  \includegraphics[width=0.9\linewidth]{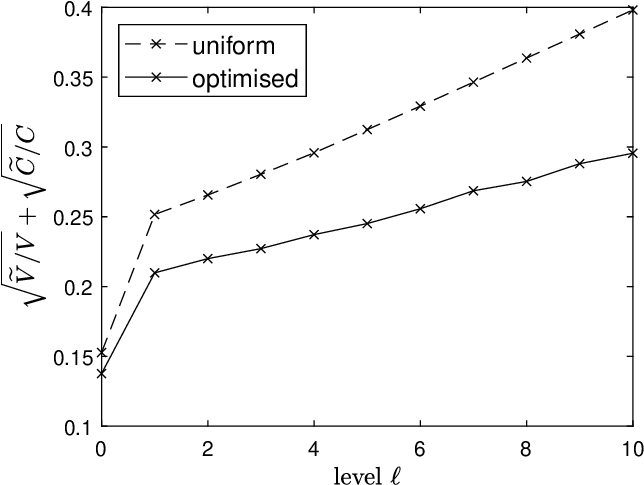}
  \captionof{figure}{$\sqrt{\Tilde{C}/C}+ \sqrt{\Tilde{V}/V}$ versus level for optimal bit-widths and best uniform bit-width.}
  \label{fig:cost_comparison}
\end{minipage}
\end{figure}

\Cref{fig:cost_lambda} shows the level cost \eqref{eq:level_cost2} versus $\lambda$ for both uniform bit-widths and optimised bit-widths. For each uniform bit-width from 4 to 16, it computes $\lambda$ based on
\begin{equation}
    \sum_i \dfrac{\partial V_\ell^{\Delta}}{\partial d_{i,\ell}} +\lambda \sum_i \dfrac{\partial \Tilde{C}_\ell}{\partial d_{i,\ell}} = 0
\end{equation}
then uses this value of $\lambda$ to determine initial values for $d_{i,\ell}$ by solving a system of equations. From the solution, it determines the values of \eqref{eq:level_cost2} for uniform and optimised bit-widths, which gives the optimal cost over $\lambda$ represented in \Cref{fig:cost_lambda}. Although we do not prove it formally we can see that the 
the resulting function 
is convex which ensures the existence of an optimum.
The optimal value for $\lambda$ is then determined by golden section search optimisation,
giving the optimal bit-widths shown in \Cref{fig:optimal_bw}, and the level cost ratios from \Cref{fig:cost_comparison}. 

The numerical results in \Cref{fig:cost_comparison} show that the cost factor for both uniform and optimised bit-widths is smaller than 1 which means that the nested framework is cheaper than the standard Multilevel Monte Carlo. \Cref{fig:cost_comparison} also confirms that the optimisation method we suggest improves the level cost compared to the best uniform bit-width choice.

The main limitation is that considering the increase of the bit widths over level, the assumption $\Tilde{C}_\ell \ll C_\ell$ might not be relevant for all levels as the cost of the fixed-point operations becomes comparable to the cost of the path generation on the CPU. 
Despite this, the nested framework is relevant at least on the first levels, which are the levels where most paths are computed, therefore the framework offers important overall savings. For instance, compared to the standard MLMC algorithm, our \Cref{fig:cost_comparison} shows a factor 7 in computational cost savings at level 0 and a factor 5 at level 1. Our experiment shows that the cost per time step and per sample on the FPGA with optimised bit-widths would be 41 for level 0 and 277 for level 1,
while the value of the cost per time step and per sample on the CPU ($=C_\ell/N$) is $C_{RNG}=10^4$.

\subsection{Rounding the bit-widths to integer values}
The next step is then rounding the bit-widths to positive integers in order to configure the fixed-point variables. In practice we observed for the single level case that the Lagrange multiplier approach was good enough compared to several integer programming approaches that we tested to optimise the bit-widths. The solution from the Lagrange multiplier approach is real and can be rounded as follows. For each level $\ell$ and each variable $i$, first round down the solution to $d^*_{i,\ell} = d_{i,\ell}$ and compute the ratio
\begin{equation}
    r_{i,\ell} = \frac{V_{indep}(d^*_{1,\ell}, \ldots, d^*_{i,\ell}, \ldots)-V_{indep}(d^*_{1,\ell}, \ldots, d^*_{i,\ell}+1, \ldots)}{\Tilde{C}(d^*_{1,\ell}, \ldots, d^*_{i,\ell}+1, \ldots)-\Tilde{C}(d^*_{1,\ell}, \ldots, d^*_{i,\ell}, \ldots)}. 
    \label{eq:ratio_lagr}
\end{equation}
Then order the ratios in decreasing order and add one bit to the variables with the highest ratio until the constraint on the error is satisfied. This heuristic performs well and is guaranteed to obtain a feasible solution.

However in the following subsection we chose to keep the bit-widths equal to the real solution in order to analyse the overall trends.

\subsection{Discussion on bit-width trends}
In order to explain the evolution of the bit-widths we use a similar approach to \cite{Rounding_error_oliver} to estimate the size of the fixed-point variables : assume that $h\ll \sqrt{h}$ and $\sigma$, $r, S_0,Z_i = \mathcal{O}(1)$, then 
we obtain
\begin{align}
    &con2, \, mul1, \, sum1,\, mul2 \sim \sqrt{h} \\
    &con1 \sim h \\
    &S \sim 1.
\end{align}
The takeaway is that the variables $S_i$ are the largest. Moreover we know that they are used in an addition at the end of each time step calculation, which explains why these variables need the largest bit-widths in order to avoid introducing extra errors and why their bit-width increases at each level as the timestep increments and the MLMC correction $\Delta P_\ell$ become smaller.

In \Cref{fig:optimal_bw}, the order and slope of the bit-widths over levels are consistent with the order of the operations in the path generation and with the size of each variable.

\begin{figure}[h]
\centering
\begin{minipage}{.48\textwidth}
  \centering
  \includegraphics[width=1\linewidth]{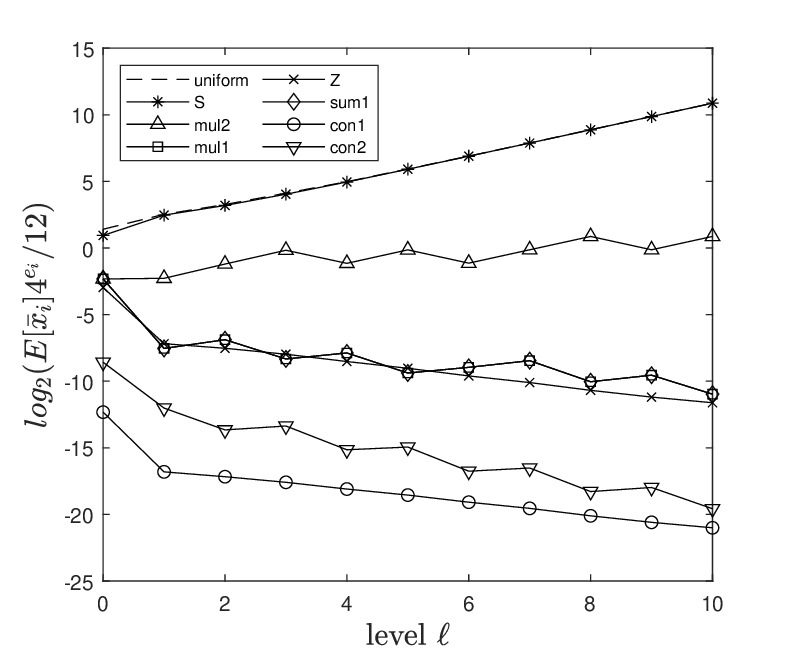}
  \captionof{figure}{Factors $\frac{1}{12}\mathbb{E}[\Bar{x}^2_i]4^{e_{i,\ell}}$ from the overall variance $V_{indep}(d)$. }
  \label{fig:var_factor}
\end{minipage}%
\hspace{0.02\textwidth}
\begin{minipage}{.48\textwidth}
  \centering
  \includegraphics[width=1\linewidth]{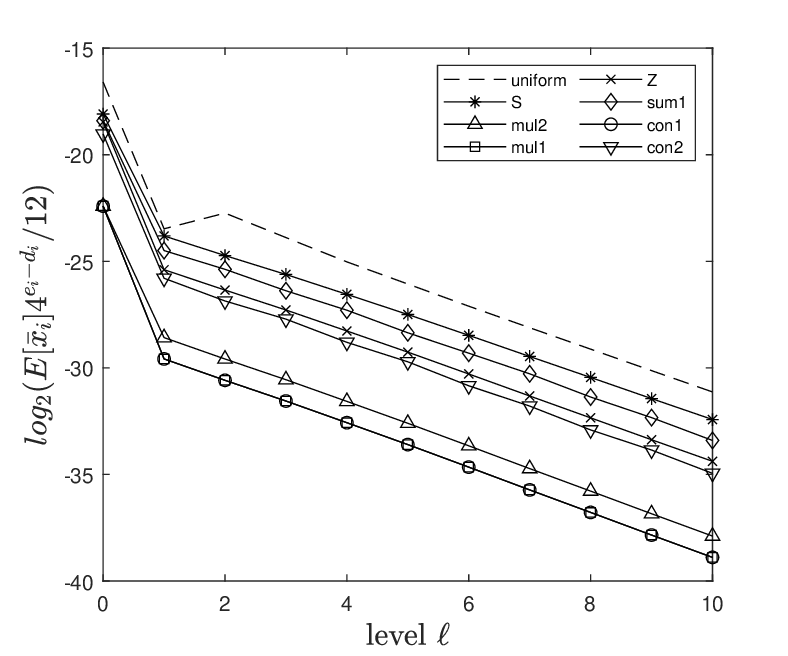}
  \captionof{figure}{Upper bounds $\frac{1}{12} \mathbb{E}[\Bar{x}^2_i]4^{e_{i,\ell}-d_{i,\ell}}$ on the expected squared errors $\mathbb{E}[\Bar{x}_i^2 \delta x_i^2]$.}
  \label{fig:indep_errors}
\end{minipage}
\end{figure}

Moreover looking at the \textit{variance factors} $\frac{1}{12}\mathbb{E}[\Bar{x}^2_i]4^{e_{i,\ell}}$ in \Cref{fig:var_factor}, for variables $S$ and $mul2$ we notice that the factors are approximately multiplied by 2 at every level. Therefore, using the fact 
% that we observe 
that $d_{S,\ell+1}=d_{S,\ell}+1$ is equivalent to 
division by 4 in the corresponding squared rounding error, the bound on the average error $\mathbb{E}[\Bar{S}_i^2 \delta S^2_i]$ is divided by 2 at every level. In fact a similar argument can be used for the error coming from $mul2$ and numerically \Cref{fig:indep_errors} confirms that all variables give a similar evolution in the squared errors. This is a very important observation because it shows that the portion of the overall error due to a certain variable is constant over levels. In other words, to leading order, all errors are roughly of the same size. This is intuitive because if an error was "disproportionately" small, there would be potential for cost savings with a small increase in the error.

As a consequence, if we sum the independent errors in \Cref{fig:indep_errors} and compare that to the Figure 4 from \cite{Rounding_error_oliver}, this shows that adapting the precision of the variables over levels allows us to keep improving the accuracy of the low precision estimate as the time step tends to 0. On the contrary, if the bit-widths in the low precision path generation were fixed, starting at a certain level, the accuracy would decrease. Indeed in the fixed precision case the rounding error at each time step of the Euler-Maruyama scheme is of order $\mathcal{O}(h^{-1}2^{e_{S,\ell}-d_{S,\ell}})$ \cite{arciniega,Rounding_error_oliver} (neglecting the effect of approximate random variables) so for large time steps the net error in $S_N$ is dominated by the time step, however for small time steps the rounding errors due to the finite precision are relatively large. Therefore %not only 
our framework not only allocates the relevant number of bits to each variable within a level but %it 
also ensures that the precision evolves with level such that the net error evolves like the time step $h$.

\section{Conclusion and future directions} \label{sec:conclusion}

In this paper we proposed a nested MLMC framework that offers important computational savings by performing most calculations in low precision and exploiting approximate random normal variables for the low precision path calculations. The low precision calculations could be performed in fixed precision on an FPGA for greater efficiency, and we suggested a procedure to optimise the bit-widths of every variable at each Monte Carlo level. This is an important improvement over previous mixed precision MLMC frameworks which held the lower precision fixed \cite{Rounding_error_oliver} or defined uniform bit-width at every level heuristically \cite{brugger2014mixed}. Our numerical results suggest that for the first levels our procedure reduces the cost at these levels by a factor 5 or 7. Hence the overall savings are significant since most paths are calculated on the first levels. Our approach would be even more efficient for the Milstein scheme because its higher order strong convergence leads to a greater proportion of the computational costs being on the coarsest levels.

The next stage of the research project will be to implement the RNG methods and the nested framework on FPGAs to determine the hardware requirements and confirm the extent of the computational savings. It would also be good to compare the performance benefits to using half-precision floating point arithmetic on GPUs or CPUs for the low-accuracy computations.

\bibliographystyle{abbrvnat}
\bibliography{main}

\begin{thebibliography}{25}
\providecommand{\natexlab}[1]{#1}
\providecommand{\url}[1]{\texttt{#1}}
\expandafter\ifx\csname urlstyle\endcsname\relax
  \providecommand{\doi}[1]{doi: #1}\else
  \providecommand{\doi}{doi: \begingroup \urlstyle{rm}\Url}\fi

\bibitem[Arciniega and Allen(2003)]{arciniega}
A.~Arciniega and E.~Allen.
\newblock Rounding error in numerical solution of stochastic differential equations.
\newblock \emph{Stoch. Anal. Appl.}, 21\penalty0 (2):\penalty0 281--300, 2003.
\newblock \doi{10.1081/SAP-120019286}.

\bibitem[Boutros et~al.(2017)Boutros, Grady, Abbas, and Chow]{HFT2}
A.~Boutros, B.~Grady, M.~Abbas, and P.~Chow.
\newblock Build fast, trade fast: {FPGA}-based high-frequency trading using high-level synthesis.
\newblock In \emph{2017 International Conference on ReConFigurable Computing and FPGAs (ReConFig)}, pages 1--6, 2017.
\newblock \doi{10.1109/RECONFIG.2017.8279781}.

\bibitem[Brugger et~al.(2014)Brugger, de~Schryver, Wehn, Omland, Hefter, Ritter, Kostiuk, and Korn]{brugger2014mixed}
C.~Brugger, C.~de~Schryver, N.~Wehn, S.~Omland, M.~Hefter, K.~Ritter, A.~Kostiuk, and R.~Korn.
\newblock Mixed precision {M}ultilevel {M}onte {C}arlo on hybrid computing systems.
\newblock In \emph{Proceedings of the IEEE Conference on Computational Intelligence for Financial Engineering \& Economics (CIFEr)}, pages 215--222. IEEE, 2014.
\newblock \doi{10.1109/CIFEr.2014.6924076}.

\bibitem[Cheung et~al.(2007)Cheung, Lee, Luk, and Villasenor]{Cheung2007HardwareGO}
R.~Cheung, D.-U. Lee, W.~Luk, and J.~Villasenor.
\newblock Hardware generation of arbitrary random number distributions from uniform distributions via the inversion method.
\newblock \emph{IEEE Transactions on Very Large Scale Integration (VLSI) Systems}, 15:\penalty0 952--962, 2007.
\newblock \doi{10.1109/TVLSI.2007.900748}.

\bibitem[Chow et~al.(2012)Chow, Tse, Jin, Luk, Leong, and Thomas]{ChowMixedPrecisionStandardMC}
G.~Chow, A.~Tse, Q.~Jin, W.~Luk, P.~Leong, and D.~Thomas.
\newblock A mixed precision {M}onte {C}arlo methodology for reconfigurable accelerator systems.
\newblock In \emph{Proceedings of the ACM/SIGDA International Symposium on Field Programmable Gate Arrays}, FPGA '12, page 57–66, New York, NY, USA, 2012. Association for Computing Machinery.
\newblock \doi{10.1145/2145694.2145705}.

\bibitem[de~Schryver et~al.(2013)de~Schryver, Torruella, and Wehn]{Schryver2013AMM}
C.~de~Schryver, P.~Torruella, and N.~Wehn.
\newblock A multi-level {M}onte {C}arlo {FPGA} accelerator for option pricing in the {H}eston model.
\newblock \emph{2013 Design, Automation \& Test in Europe Conference \& Exhibition (DATE)}, pages 248--253, 2013.
\newblock \doi{10.7873/DATE.2013.063}.

\bibitem[Gaffar et~al.(2002)Gaffar, Mencer, Luk, Cheung, and Shirazi]{bitwidth-AD}
A.~Gaffar, O.~Mencer, W.~Luk, P.~Cheung, and N.~Shirazi.
\newblock Floating-point bitwidth analysis via automatic differentiation.
\newblock In \emph{2002 IEEE International Conference on Field-Programmable Technology}, pages 79--88, 2002.
\newblock \doi{10.1109/FPT.2002.1188677}.

\bibitem[Gaffar et~al.(2004)Gaffar, Mencer, Luk, and Cheung]{unifying-bwoptim}
A.~Gaffar, O.~Mencer, W.~Luk, and P.~Cheung.
\newblock Unifying bit-width optimisation for fixed-point and floating-point designs.
\newblock In \emph{12th Annual IEEE Symposium on Field-Programmable Custom Computing Machines}, pages 79--88, 2004.
\newblock \doi{10.5555/1025123.1025818}.

\bibitem[Giles(2008)]{GILES2008}
M.~Giles.
\newblock Multilevel {M}onte {C}arlo path simulation.
\newblock \emph{Oper. Res.}, 56\penalty0 (3):\penalty0 607--617, 2008.
\newblock \doi{10.1287/opre.1070.0496}.

\bibitem[Giles(2015)]{Giles_overview17}
M.~Giles.
\newblock Multilevel {M}onte {C}arlo methods.
\newblock \emph{Acta Numer.}, 24:\penalty0 259–328, 2015.
\newblock \doi{10.1017/S096249291500001X}.

\bibitem[Giles and Sheridan-Methven(2021)]{NestedOliver}
M.~Giles and O.~Sheridan-Methven.
\newblock Analysis of nested multilevel {M}onte {C}arlo using approximate normal random variables.
\newblock \emph{SIAM/ASA J. Uncertain. Quantif.}, 10:\penalty0 200--226, 2021.
\newblock \doi{10.1137/21M1399385}.

\bibitem[Giles and Sheridan-Methven(2023)]{approximateICDF_Oliver}
M.~Giles and O.~Sheridan-Methven.
\newblock Approximating inverse cumulative distribution functions to produce approximate random variables.
\newblock \emph{ACM Trans. Math. Software}, 49\penalty0 (3), sep 2023.
\newblock \doi{10.1145/3604935}.

\bibitem[Glasserman(2004)]{glasserman04}
P.~Glasserman.
\newblock \emph{{M}onte {C}arlo Methods in Financial Engineering}.
\newblock Springer, New York, 2004.
\newblock \doi{10.1007/978-0-387-21617-1}.

\bibitem[Intel()]{norminv_routine_intel}
Intel.
\newblock Developer reference for {I}ntel® one{API} {M}ath {K}ernel {L}ibrary for {C}. {v?CdfNormInv}.
\newblock Available at : \url{https://www.intel.com/content/www/us/en/docs/onemkl/developer-reference-c/2024-1/v-cdfnorminv.html}.
\newblock (Accessed: 21 November 2024).

\bibitem[Leber et~al.(2011)Leber, Geib, and Litz]{HFT1}
C.~Leber, B.~Geib, and H.~Litz.
\newblock High frequency trading acceleration using {FPGA}s.
\newblock In \emph{2011 21st International Conference on Field Programmable Logic and Applications}, pages 317--322, 2011.
\newblock \doi{10.1109/FPL.2011.64}.

\bibitem[Lee et~al.(2004)Lee, Luk, Villasenor, and Cheung]{thom14}
D.-U. Lee, W.~Luk, J.~D. Villasenor, and P.~Y.~K. Cheung.
\newblock A {G}aussian noise generator for hardware-based simulations.
\newblock \emph{IEEE Trans. Comput.}, 53\penalty0 (12):\penalty0 1523–1534, Dec 2004.
\newblock \doi{10.1109/TC.2004.106}.

\bibitem[Lee et~al.(2006)Lee, Gaffar, Cheung, Mencer, Luk, and Constantinides]{Lee2006}
D.-U. Lee, A.~Gaffar, R.~Cheung, O.~Mencer, W.~Luk, and G.~Constantinides.
\newblock Accuracy-guaranteed bit-width optimization.
\newblock \emph{IEEE Transactions on Computer-Aided Design of Integrated Circuits and Systems}, 25\penalty0 (10):\penalty0 1990--2000, October 2006.
\newblock \doi{10.1109/TCAD.2006.873887}.

\bibitem[Lee et~al.(2009)Lee, Cheung, Luk, and Villasenor]{Lee_segmentation}
D.-U. Lee, R.~Cheung, W.~Luk, and J.~Villasenor.
\newblock Hierarchical segmentation for hardware function evaluation.
\newblock \emph{IEEE Transactions on Very Large Scale Integration (VLSI) Systems}, 17\penalty0 (1):\penalty0 103--116, 2009.
\newblock \doi{10.1109/TVLSI.2008.2003165}.

\bibitem[Lindsey et~al.(2016)Lindsey, Leslie, and Luk]{lindsey2016domain}
B.~Lindsey, M.~Leslie, and W.~Luk.
\newblock A {D}omain {S}pecific {L}anguage for accelerated {M}ultilevel {M}onte {C}arlo simulations.
\newblock In \emph{Proceedings of the IEEE 27th International Conference on Application-specific Systems, Architectures and Processors (ASAP)}. IEEE, 2016.
\newblock \doi{10.1109/ASAP.2016.7760778}.

\bibitem[Malik and Hemani(2016)]{Malik2016GaussianRN}
J.~S. Malik and A.~Hemani.
\newblock Gaussian random number generation: A survey on hardware architectures.
\newblock \emph{ACM Comput. Surv.}, 49\penalty0 (3), Nov. 2016.
\newblock \doi{10.1145/2980052}.

\bibitem[MathWorks()]{fi_toolbox}
MathWorks.
\newblock Fixed-{P}oint {D}esigner documentation.
\newblock Available at : \url{https://uk.mathworks.com/help/fixedpoint/}.
\newblock (Accessed: November 2024).

\bibitem[Menon et~al.(2018)Menon, Lam, Osei-Kuffuor, Schordan, Lloyd, Mohror, and Hittinger]{ADAPT}
H.~Menon, M.~Lam, D.~Osei-Kuffuor, M.~Schordan, S.~Lloyd, K.~Mohror, and J.~Hittinger.
\newblock {ADAPT}: Algorithmic differentiation applied to floating-point precision tuning.
\newblock In \emph{SC18: International Conference for High Performance Computing, Networking, Storage and Analysis}, pages 614--626, 2018.
\newblock \doi{10.1109/SC.2018.00051}.

\bibitem[Sheridan-Methven and Giles(2024)]{Rounding_error_oliver}
O.~Sheridan-Methven and M.~Giles.
\newblock Rounding error using low precision approximate random variables.
\newblock \emph{SIAM J. Sci. Comput.}, 46\penalty0 (4), 2024.
\newblock \doi{10.1137/23M1552814}.

\bibitem[Thomas et~al.(2009)Thomas, Howes, and Luk]{Thomas2009ACO}
D.~Thomas, L.~Howes, and W.~Luk.
\newblock A comparison of {CPU}s, {GPU}s, {FPGA}s, and massively parallel processor arrays for random number generation.
\newblock In \emph{Symposium on Field Programmable Gate Arrays}, 2009.
\newblock \doi{10.1145/1508128.1508139}.

\bibitem[Woods et~al.(2008)Woods, McAllister, Lightbody, and Yi]{woods2008fpga}
R.~Woods, J.~McAllister, G.~Lightbody, and Y.~Yi.
\newblock \emph{{FPGA}-based implementation of signal processing systems}.
\newblock John Wiley \& Sons, 2008.
\newblock \doi{10.1002/9781119079231}.

\end{thebibliography}

\end{document}